\documentclass[floatfix,aps,twocolumn,preprintnumbers,amsmath,amssymb,nofootinbib,superscriptaddress]{revtex4}

\usepackage{graphicx,color}

\usepackage[final]{changes}
\definecolor{addedcolor}{rgb}{0,0,1} 
\definecolor{deletedcolor}{rgb}{1,0,0} 
\setdeletedmarkup{\color{deletedcolor}\sout{#1}}
\setaddedmarkup{\textcolor{addedcolor}{#1}}

\def\1{\'{\i}}

\def\XXint#1#2#3{{\setbox0=\hbox{$#1{#2#3}{\int}$}
     \vcenter{\hbox{$#2#3$}}\kern-.5\wd0}}

\newcommand{\nh}{{\bf h}}

\newcommand{\nk}{{\bf k}}
\newcommand{\np}{{\bf p}}
\newcommand{\nq}{{\bf q}}

\def\XXint#1#2#3{{\setbox0=\hbox{$#1{#2#3}{\int}$}
     \vcenter{\hbox{$#2#3$}}\kern-.5\wd0}}

\def\1{\'{\i}}

\begin{document}


\title{
Charged-current quasielastic neutrino scattering from $^{12}$C in an 
extended superscaling model with two-nucleon emission 
}


\author{V.L. Martinez-Consentino}
\email{victormc@ugr.es}
\affiliation{Departamento de
  F\'{\i}sica At\'omica, Molecular y Nuclear,
Universidad de Granada,  E-18071 Granada, Spain.}
\affiliation{ Instituto Carlos I
  de F{\'\i}sica Te\'orica y Computacional, Granada, Spain.}

\author{J.E. Amaro}
\email{amaro@ugr.es} 
\affiliation{Departamento de
  F\'{\i}sica At\'omica, Molecular y Nuclear,
Universidad de Granada,  E-18071 Granada, Spain.}
\affiliation{ Instituto Carlos I
  de F{\'\i}sica Te\'orica y Computacional,  Granada, Spain.}


\begin{abstract}

The quasielastic cross-section of charged-current neutrino and
antineutrino scattering on $^{12}$C is calculated using an improved
superscaling model with relativistic effective mass.  Our model
encompasses two-particle emission induced by neutrinos, which we
distinguish into two contributions. The first contribution arises from
meson-exchange currents, and its calculation is performed at a
microscopic level. The second contribution is phenomenological and
extracted from the high-energy tail of the scaling function, assumed
to be produced by 2p2h mechanisms where the one-body current plays a role,
such as short-range correlations and interferences with MEC,
final-state interaction, etc.
The model explicitly includes the modification of the
relativistic effective mass of the nucleon within the relativistic
mean field model of nuclear matter. The meson exchange currents are
also consistently calculated within the same model. 
 With this model, we present predictions
for the neutrino and antineutrino cross sections of $^{12}$C that have
been measured in accelerator experiments.

\end{abstract}


\maketitle

\section{Introduction}

Neutrino interactions with nuclei are of vital importance for a series
of experiments ongoing and planned in the near future \cite{Bal22}. In
recent years, a large experimental program focused on
accelerator-based neutrino oscillation experiments has been
developed. The primary objective of this program is to enhance our
understanding of neutrino properties by accurately measuring the
oscillation parameters and investigating the weak CP-violating phase.
These measurements are crucial for elucidating the enigmatic phenomena
surrounding neutrino oscillations and their fundamental role in the
realm of particle physics.  The majority of long-baseline neutrino
experiments employ complex nuclear targets, making it crucial to have
excellent control over medium effects in neutrino-nucleus scattering
to achieve precise measurements of neutrino oscillation parameters
\cite{Abe20,Mos16,Kat17}.

The cross section of neutrino-nucleus scattering for the energies of
interest, which are essential for these experiments \cite{Alv18,Ank22}, is
made up of various channels, including quasielastic interactions.  But
the nuclear models present uncertainties \cite{Rut19,Pat18} related
to the presence of processes, such as final-state interactions,
multi-nucleon emission, pion emission and other inelasticities that
limit the accuracy of oscillation experiments
\cite{Abe11,Abe18a,Abe18b,Ada16,Acc15}. Efforts to reduce
uncertainties in neutrino-nucleus scattering models include measuring
the proton or neutron multiplicity of final states \cite{Abe18b,
  Pal16, Acc14,Bac17}.

The inclusive $(e,e')$ cross sections of nuclei provide a useful
starting point for modeling neutrino-nucleus interactions. The
isovector component of the electromagnetic nuclear responses can be
related to the vector component of the weak charged-current responses
that contribute to the neutrino cross sections. 
Therefore, the systematic differences observed
between theoretical predictions of neutrino cross sections should be
closely associated with differences in the description of $(e,e')$
data \cite{Mar09,Nie11,Gal16,Meg16,Roc16,Mar16}.

In this work, we present predictions for neutrino quasielastic cross
sections based on an extended superscaling model that has been tuned from
inclusive electron scattering data. It is based on the superscaling
analysis with relativistic effective Mass (SuSAM*) model of
refs. \cite{Ama15,Ama17}, modified to accommodate the possibility of a
contribution that partially violates scaling.  This modification
acknowledges that the inclusive Quasielastic (QE) cross section data exhibit
deviations from strict scaling behavior.

The original SuSAM* model differs from the superscaling analysis
(SuSA) model \cite{Alb88,Day90,Don99,Ama21} in that it is based on the
relativistic mean-field model \added{(RMF)} of nuclear matter and thus contains, by
construction, nucleon medium effects encoded in a relativistic
effective mass for the nucleon.

Our modified SuSAM*+2p2h model is made up of three contributions: i) the purely
quasielastic cross section (1p1h), ii) the two-particle two-hole
(2p2h) emission channel from a theoretical model of meson-exchange
currents (MEC), and iii) an additional phenomenological 2p2h tail
contribution attributed to short-range correlations (SRC),
interference with MEC, final-state
interactions (FSI) and other processes partially contributing to the
scaling function, exhibiting slight deviations from scaling behavior
and mainly influencing the high-energy tail of the scaling function.

The 2p2h-MEC responses are calculated consistently with the mean field
model in nuclear matter by introducing effective mass and vector
energy for the nucleon, so they explicitly contain the same medium
corrections as the quasielastic responses \cite{Mar21,Mar21b}.  The
contribution of the 2p2h tail has been obtained phenomenologically
from $(e,e')$ data through a pure phase-space model \cite{Mar23}
similar to the one implemented in the GiBUU event generator
\cite{Mos14}, by a parameterization involving the phase-space function
of the 2p2h process and the single-nucleon responses multiplied by a
$q$-dependent 2p2h parameter.  An additional improvement is that the
single nucleon responses that are factorized in the scaling approach
consist in an average with respect to a modified momentum distribution instead
of extrapolations from the relativistic Fermi gas (RFG) model
\cite{Mar23,Cas23,Cas23b}.

 The SuSAM*+2p2h model was fitted to quasielastic  electron \deleted{quasielastic}
 data in Ref. \cite{Mar23} 
and therefore reproduces a large part of them (excluding
 inelastic and pion emission data). Here we intend to apply it to calculate
 the neutrino cross section to see what is achieved, as was done with
 the original SuSA model \cite{Ama05a,Ama05b}.  The SuSAM* model is
 based on the relativistic mean field model \cite{Ros80,Ser86}, and it
 provides an interesting alternative to the traditional SuSA model
 \cite{Meg16,Meg18}.  Indeed, the relativistic mean field is an
 optimal starting point for modeling the nuclear response, as it
 already adequately reproduces the position and width of the
 quasielastic peak with only two parameters, the Fermi momentum and
 the effective mass \cite{Ros80}. Besides it is relativistic, gauge
 invariant, and implicitly contains dynamical enhancement of lower
 Dirac components of the nucleon in the medium.

The structure of this work is as follows. In Sect. II we briefly describe the
theoretical SuSAM*+2p2h model. In Sect. III we provide the
results for the neutrino and antineutrino cross sections and
and compare with  data. In sect. IV we give our summary and conclusions.

\section{Formalism}

\subsection{Response functions}

In this work, our primary focus lies on studying inclusive
charged-current quasielastic scattering (CCQE) of muon neutrinos from
nuclei.  In this section, we provide a concise overview of the
formalism, which is elaborated in more detail in refs. \cite{Ama20,Mar23}.
In the case of $(\nu_{\mu},\mu^-)$ 
reaction the initial neutrino energy is denoted as $\epsilon=E_\nu$, and the
final detected muon energy is represented as $\epsilon'=m_\mu +T_\mu$,
where $m_\mu$ denotes the \deleted{muon} mass \added{ and $T_\mu$ is the kinetic energy of the muon.}.  The four-momentum transfer is
defined as $Q^\mu=k^\mu-k'{}^\mu=(\omega,\nq)$, with $Q^2=\omega^2-q^2
< 0$, and the initial and final lepton momenta are $k^\mu$ and
$k'{}^\mu$.   The scattering angle $\theta$ is given by $\nk\cdot\nk'=
kk'\cos\theta$.

In this paper, our interest lies in the inclusive cross section, where
only the final muon is detected.  The cross section is expressed as
follows:
\begin{equation}
\frac{d^2\sigma}{d\cos\theta dE_\mu} = \frac{G_F^2\cos^2\theta_c}{2\pi} 
\frac{k'}{\epsilon} L_{\mu\nu}W^{\mu\nu},
\end{equation}
where \(G_F\) is the Fermi constant, \(\theta_c\) is the Cabibbo
angle, 
and \(L_{\mu\nu}\)
and \(W^{\mu\nu}\) represent the leptonic and hadronic tensors,
respectively. 
The leptonic tensor for neutrino scattering is given by:
\begin{equation} L_{\mu\nu} = 
 k_\mu k'_\nu + k'_\mu k_\nu - g_{\mu\nu} k \cdot k'
\pm i\epsilon_{\mu\nu\alpha\beta} k^{\alpha}k'{}^\beta, 
\end{equation}
where the sign $+$ $(-)$ is for neutrino (antineutrino) scattering. 
The inclusive hadronic tensor is constructed from the matrix elements
of the current operator $J^\mu(Q)$ between the initial and final
hadronic states, summing over all the possible final nuclear states
with excitation energy $\omega=E_f-E_i$, and averaging over the
initial spin components.
\begin{equation}
W^{\mu\nu}
= 
\sum_f \overline{\sum_i} 
\langle f | J^\mu(Q) |i \rangle^*
\langle f | J^\nu(Q) |i \rangle
\delta(E_i+\omega-E_f) .
\label{hadronic-tensor}
\end{equation}
To express the
cross section in a simple form in terms of response functions, we
utilize the $q$-reference system where the momentum transfer $\nq$
aligns with the $z$-axis, and the $x$-axis corresponds to the component
perpendicular to $q$ of the incident neutrino momentum.  In this 
$q$-system, only the following five components of the hadronic
tensor or response functions are involved:
\begin{eqnarray}
R^{CC} &=&  W^{00} \label{rcc} \\
R^{CL} &=& -\frac12\left(W^{03}+ W^{30}\right) \\
R^{LL} &=& W^{33}  \\
R^{T} &=& W^{11}+ W^{22} \\
R^{T'} &=& -\frac{i}{2}\left(W^{12}- W^{21}\right)\,. \label{rtprima}
\end{eqnarray}

\subsection{Relativistic mean field}

The SuSAM*+2p2h model considered in this work is based on the relativistic
mean field \deleted{(RMF)} model of nuclear matter, as initially proposed in the
references \cite{Ros80, Ser86, Weh93, Bar98}. In this model, nucleons
interact with a relativistic field characterized by scalar and vector
potentials.  The single-particle wave functions are
described by plane waves with momentum \( \mathbf{p} \), and their
on-shell energy is given by:
\begin{equation}
 E = \sqrt{(m_N^*)^2 + \mathbf{p}^2}, 
\end{equation}
where \( m_N^* \) represents the relativistic effective mass of the
nucleon, defined as $ m_N^* = m_N - g_s\phi_0 = M^*m_N$.  Here, \( m_N
\) denotes the bare nucleon mass, and \( g_s\phi_0 \) represents the
scalar potential energy of the RMF \cite{Ser86}. \deleted{ For the nucleus
\(^{12}\)C considered in our results, the value of \( M^* \) is chosen
to be 0.8 }.  Additionally, due to the repulsion by the relativistic
vector potential, the nucleon acquires a positive energy given by \(
E_v = g_vV_0 \).  Hence, the total nucleon energy in the RMF model is
the sum of the on-shell energy and the vector potential energy,
$E_{RMF} = E + E_v$. According to ref. \cite{Mar21} 
we use the values
$M^* = 0.8$ and $E_v=141$ MeV for \(^{12}\)C in our results.

In the RMF \added{framework}, the
generic quasielastic responses with a one-body current operator can
be expressed as follows:
\begin{equation} \label{RFG}
R_{QE}^K(q,\omega)
= \frac{\epsilon_F-1}{m_N^*\eta_F^3\kappa} {\cal N} \overline{U}_K(q,\omega) f^*(\psi^*)
\end{equation}
where ${\cal N}=Z, N$ for protons or neutrons responses,
$\kappa=q/2m_N^*$ \deleted{, and $\epsilon_F=E_F/m_N^*$} and
$\eta_F=k_F/m_N^*$ are the Fermi energy and momentum, respectively, in
units of the effective mass, with $E_F=\sqrt{k_F^2+m_N^{*2}}$.  The
scaling variable $\psi^*=\psi^*(q,\omega)$ is related to the minimum
energy of an on-shell nucleon absorbing momentum
$q$ and energy $\omega$,
\begin{equation}
\psi^* = \sqrt{\frac{\epsilon_0-1}{\epsilon_F-1}} {\rm sgn} (\lambda-\tau),
\end{equation}
where $\epsilon_F=E_F/m_N^*$ and
\begin{equation}
\epsilon_0={\rm Max}
\left\{ 
       \kappa\sqrt{1+\frac{1}{\tau}}-\lambda, \epsilon_F-2\lambda
\right\},
\end{equation}
\deleted{where} we use the usual
dimensionless variables normalized with $m_N^*$ 
\begin{equation}
\lambda  = \omega/2m_N^*, \kern 1cm
\tau  =  \kappa^2-\lambda^2. 
\end{equation}
Finally, the single nucleon functions 
$\overline{U}_K$  represent the
responses of a single nucleon averaged in the Fermi gas \cite{Ama20}.

In the RMF the scaling function is
\begin{equation}
f^*(\psi^*)=\frac34(1-\psi^{*2})\theta(1-\psi^{*2}),
\end{equation}
 and therefore is
zero outside the Fermi gas region 
$-1<\psi^* <1$.  This is a consequence
of the fact that nucleons have a maximum momentum \(
k_F \), which implies that, for a fixed \( q \), there exists a
maximum and minimum energy transfer \deleted{(\( \omega \))} where the response
is non-zero. 

\subsection{Extended superscaling}

In a finite nuclear system like an actual nucleus, the quasielastic
response extends beyond the region $-1<\psi^* <1$.  In the SuSAM*
approach, the factorization, Eq. (10), remains intact, but using a
phenomenological scaling function that is obtained from experimental
data.  As a result, the SuSAM* model can account for the complexities
arising from finite nuclear systems and better describe the cross
section.  This extension also necessitates broadening the definition
of the averaged single-nucleon responses beyond the RFG region, which
can be achieved through a smearing of the Fermi surface, as
implemented in \cite{Cas23,Cas23b}.  In this context, we perform an averaging
of the single-nucleon using the momentum distribution that is deduced
from the scaling function. This procedure is elaborated upon in
Appendix B of the reference \cite{Mar23}.

The first step in the scaling analysis involves subtracting the
contribution of 2p2h-MEC responses from the $(e, e')$ data.  The MEC
responses are calculated in the RMF model of Ref. \cite{Mar21}.  
The experimental data of the scaling
function \added{,}\( f^*_{\rm exp} \)\added{,} are then obtained by
dividing by the averaged single-nucleon cross section. 
\begin{equation}
f^*_{\rm exp} =\frac{ (\frac{d\sigma}{d\Omega d\omega})_{exp}
  -(\frac{d\sigma}{d\Omega d\omega})_{MEC}
}{\sigma_M ( v_L r_L + v_T
  r_T) },
\label{fexp}
\end{equation}
where $\sigma_M$ is the Mott cross section, and 
the single nucleon dividing responses are such that the cancel with the 
factor multiplying the scaling function in Eq. (10) 
\begin{equation} \label{rsn}
r_K = \frac{\epsilon_F-1}{m_N^* \eta_F^3 \kappa} 
(Z \overline{U}^p_K+N \overline{U}^n_K).
\end{equation}
In this way the contamination arising from 2p2h-MEC response has been
minimized to the greatest extent possible within the scaling data.

The process of selecting the QE-like data is essential for the SuSAM*
approach, and it involves representing the $f^*_{\rm exp}$ data as a
function of the scaling variable $\psi^*$. The data points belonging
to the QE peak are expected to exhibit approximate scaling
when plotted against $\psi^*$. While they may not align perfectly with
the scaling function of the RFG, they predominantly cluster around a
thick band near the RFG. On the other hand, data points corresponding
to non-quasielastic processes like pion production clearly deviate
from this band, scattering them in disparate points. This distinction
enables the selection of QE-like data and the elimination of
non-scaling inelastic data (approximately half of the $^{12}$C
dataset). The data selection process is formalized using a density
algorithm: a data point is selected if it is surrounded by more than
25 points within a radius $r= 0.1$. The density algorithm automatically
identifies points that collapse into a dense region (scaling) and
removes dispersed data points that do not collapse (non-scaling).

The selected QE-like data were used in reference \cite{Mar21} to
construct a phenomenological SuSAM* scaling function, $f^*(\psi^*)$,
which is parameterized as the sum of two Gaussians and exhibits a
clear asymmetry around the origin $\psi^*=0$, with a tail for
$\psi^*>0$.  In reference \cite{Mar23}, it was noticed that this tail
is consistent with a contribution from 2p-2h processes 
involving the one-body (OB) current. A significant portion of these
effects could potentially originate from SRC that generate high-momentum nucleons in the ground state.  
 Therefore, in the extended SuSAM* approach (ESuSAM*) an alternative
parametrization of the phenomenological scaling function is proposed
in the form
\begin{equation} \label{extended}
f^*_{E}(q,\omega) = f^*_{1p1h}(\psi^*)+ f^*_{SRC}(q,\omega)
\end{equation}
where $f^*_{1p1h}(\psi^*)$ is a symmetric function that can be 
parametrized with a  Gaussian
\begin{equation}\label{f1p1h}
f_{1p1h}^* (\psi^*) = b e^{-(\psi^*)^2 / a^2}.
\end{equation}
with the coefficients a=$0.744$ and $ b=0.682$.  The function
$f^*_{SRC}(q,\omega)$ characterizes the tail behavior of the scaling
data for $\psi^*>0$. It is formulated as a pure phase-space model (as
shown below) because it is assumed to represent
2p2h contributions involving the OB current. This primarily
includes  SRC,
interferences with MEC, and possibly other processes
such as FSI. In other words, it cannot be solely
attributed to SRC effects, even though we label it as such to
differentiate it from the pure 2p2h-MEC
contribution.

\begin{figure}[ht]
\centering
\includegraphics[width=8cm]{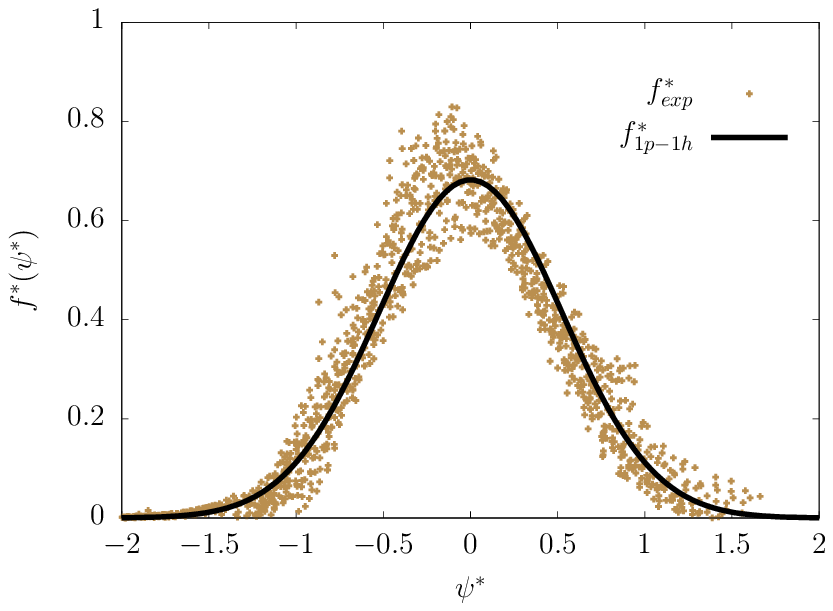}
\caption{
The phenomenological scaling function used in this article compared to experimental data after subtracting the theoretical contribution of 2p2h. 
}\label{fig1}
\end{figure}

In Figure 1, we present the experimental QE scaling function data
after subtracting the contribution of the tail,
$f^*_{SRC}(q,\omega)$. It is evident that the data now conforms to a
band that aligns with a symmetric distribution, fitting well with a
Gaussian centered at \(\psi^*=0\). This observation implies that the
kinematic dependence of the tail, as modeled by the pure phase-space
approach, is suitable for describing the high-energy tail accurately.
The width of the band provides an indication of the extent to which
the scaling hypothesis is violated.

Inserting the extended scaling function $f^*_E(q,\omega)$ 
in Eq. (\ref{RFG}),
the QE responses \deleted{(\ref{RFG})} are the sum of two contributions
\begin{equation}
R^K_{QE}(q,\omega) = R^K_{1p1h}(q,\omega)+R^K_{SRC}(q,\omega)
\end{equation}
where $R^K_{1p1h}$ is proportional to $f^*_{1p1h}(\psi^*)$ and
$R^K_{SRC}$ is proportional to $f^*_{SRC}(q,\omega)$ using Eq. (10). 
Thus,
the former scales and the latter does not.  However, in this model,
both responses are proportional to the averaged single-nucleon
response. 
The cross sections obtained using this ESuSAM* approach 
closely resemble those obtained with the original SuSAM* model,
as both models have been fitted to the same dataset. However, the
advantage of the extended model lies in its explicit ability to
separate the contribution of the high-energy tail from the symmetric
part of the scaling function, which lacks a tail. This feature becomes
particularly useful in applications such as neutrino event generators,
where distinguishing between different types of final states is
essential. In the case of the \(R_{1p1h}\) response, the final states
would be genuine 1p1h excitations, while for the \(R_{SRC}\) response,
these states correspond to 2p2h excitations.

Finally the total responses in the ESuSAM* + MEC model 
are written as the sum of three contributions
\begin{equation}
R^K = R^K_{1p1h}+R^K_{SRC}+R^K_{MEC}.
\end{equation}
The MEC contributions need to be added back because they were 
previously subtracted from the data to obtain
 the phenomenological scaling function $f^*(q,\omega)$.
Note that pion emission and other inelastic channels
are not included in this model.

\subsection{Phase-space model}

Next, we proceed to describe the pure phase space model for the SRC 
responses that contribute to the tail of the scaling function.  The equations were derived in Ref. \cite{Mar23}
from a factorized model for
 the emission of two correlated nucleons within
the framework of the independent particle approximation. Subsequently,
this factorized approximation was fitted to describe the tail of the
phenomenological scaling function. 
The corresponding response functions
for symmetric nuclei ($N=Z$)  are the following
\begin{equation} \label{SRC}
R_{SRC}^K= \frac{V F(q,\omega)}{(2\pi)^9}
\frac{Z+\alpha(Z-1)}{2Z-1}
\frac{c^{pn}(q)}{m_N^2m_\pi^4}
\overline{U}^K
\end{equation}
where $V/(2\pi)^3 = Z/(\frac8 3 \pi k_F^3)$. 

In Eq. (\ref{SRC}) the function $F(q,\omega)$ is the phase space function 
for the 2p2h excitations in the RMF model of nuclear matter
 \begin{eqnarray}
F(q,\omega)
&=& \int
d^3p'_1
d^3p'_2
d^3h_1
d^3h_2
\frac{(m^*_N)^4}{E_1E_2E'_1E'_2}
\nonumber\\
&&
\theta(p'_1-k_F)
\theta(k_F-h_1) 
\theta(p'_2-k_F)
\theta(k_F-h_2) 
\nonumber\\
&&
 \times
 \delta(E'_1+E'_2-E_1-E_2-\omega)
\nonumber\\
&&
\times 
\delta(\np'_1+\np'_2-\nq-\nh_1-\nh_2) 
\end{eqnarray}
In Eq. (\ref{SRC}) the parameter $\alpha=1/18$ take into account the
dominance $np$ pairs over $pp$ pairs in the high-momentum
distribution, while the factor $[Z+\alpha(Z-1)]/(2Z-1)$
takes into account the percent contribution of pn, pp and nn
pairs. 

Finally \added{,} the coefficients $c^{pn}(q)$ are the 2p2h parameters.
 The dividing factor $m_N^2m_\pi^4$ is introduced for convenience
so that the $c^{pn}$ coefficients are adimensional. In the factorized
independent-pair approximation the $c^{pn}$ coefficients are
proportional to the average value of the high momentum distribution of
a pn pair in the 2p2h excitation. In ref. \cite{Mar23} the 
2p2h parameters were
fitted  to describe the tail of the scaling function for each value of
$q$. The $q$-dependence is well described by the approximate formula
\begin{equation}
c^{pn}(q)=a_0 \sqrt{\frac{m_N^*+q}{q}}\frac{m_N^*}{q}
\end{equation}
with $a_0=345$. 

The SRC response functions, Eq. (\ref{SRC}) can be analytically
evaluated using the frozen nucleon approximation in the 
phase-space function that can be approximated by a simple formula
\cite{Rui14}:
\begin{equation}
F(q,\omega)\simeq 
4\pi
\left( \frac{4}{3}\pi k_F^3 \right)^2
\frac{m_N^{* 2}}{2} \sqrt{1-\frac{4m_N^{*2}}{(2m_N^*+\omega )^2-q^2}}\,,
\label{analytical}
\end{equation}
This formula was utilized in Ref. \cite{Mar23} in the fit of the 2p2h
coefficients.
From this equation 
we obtain the minimum $\omega$  to excite a 2p2h state
 for fixed $q$, in frozen approximation, corresponding to 
the solution of $F(q,\omega)=0$ 
\begin{equation}
 \omega_{\rm min} =  \sqrt{4 m_N^{*2} + q^2} - 2 m^*_N
    \label{correctionf}
\end{equation}
The phase-space function is set to zero below this value. Note that this is
the kinetic energy of a particle with mass $2m_N^*$ and momentum $q$.

The factorization of Eq. (\ref{SRC}) is similar to the pure phase-space
model proposed in reference \cite{Lal12}, with the distinction that
our model incorporates an additional $q$-dependence in the 2p2h
parameters. This extra $q$-dependence is necessary to accurately depict
the tail of the scaling data in the quasielastic region \cite{Mar23}. 

\subsection{Meson-exchange currents}

\begin{figure}[t]
\centering
\includegraphics[width=8cm]{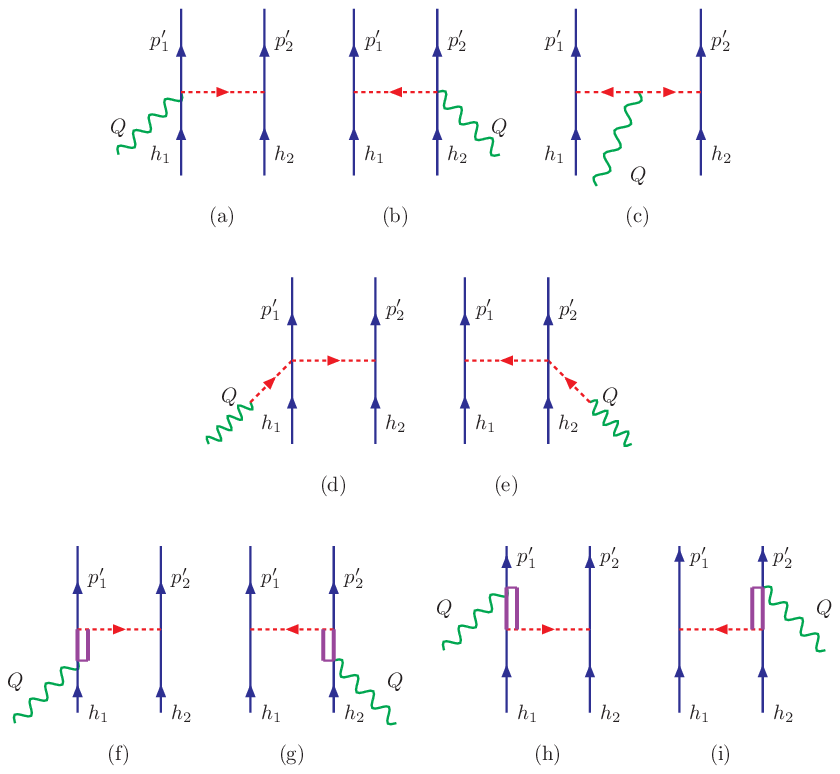}
\caption{Feynman diagrams for the electroweak MEC model used in 
this work.}\label{fig-feynman}
\end{figure}

The relativistic MEC model is adopted from Refs.
\cite{Rui17,Mar21}, which builds upon the 
weak pion production model of Ref. \cite{Her07}\added{.}
This model encompasses the Feynman
diagrams depicted in Figure \ref{fig-feynman}, 
categorized as seagull (a,b), pion-in-flight (c), 
pion-pole (d,e), and $\Delta$ (f--i)  currents. Notably, the 
$ \Delta$ current
contribution proves to be the most significant within the momentum
range of $q=500$ -- 1000 MeV/c, which holds utmost relevance for
neutrino scattering.

The matrix elements of the MEC operator between the ground state and 
2p2h excitations in the Fermi gas are given by
\begin{eqnarray}
\langle p'_1p'_2h_1^{-1}h_2^{-1}| J^\mu(Q)| F \rangle
&=&
\frac{(2\pi)^3}{V^2}
\frac{(m^*_N)^2}{\sqrt{E_1E_2E'_1E'_2}}
\nonumber\\
&&
\kern -4cm
\times 
\delta(\np'_1+\np'_2-\nq-\nh_1-\nh_2)
j^{\mu}(p'_1,p'_2,h_1,h_2),
\label{MEC}
\end{eqnarray}
 where the function $j^{\mu}(p'_1,p'_2,h_1,h_2)$ incorporates the spin
 and isospin indices of the Dirac spinors in the 2p2h excitation. The
 explicit expressions for these functions can be found in
 Ref. \cite{Mar21}.  The 2p2h-MEC responses are computed here within
 the framework of the RMF theory for nuclear
 matter. Consequently, the Dirac spinors are associated with a
 relativistic effective mass \(m_N^*\).
The corresponding 2p2h hadronic tensor 
is computed as 
 \begin{eqnarray}
W^{\mu\nu}_{\rm MEC}
&& 
=\frac{V}{(2\pi)^9}\int
d^3p'_1
d^3p'_2
d^3h_1
d^3h_2
\frac{(m^*_N)^4}{E_1E_2E'_1E'_2}
\\
&&
 \times  w^{\mu\nu}(\np'_1,\np'_2,\nh_1,\nh_2) \delta(E'_1+E'_2-E_1-E_2-\omega)
\nonumber\\
&&
\times 
\theta(p'_1-k_F)
\theta(k_F-h_1) 
\theta(p'_2-k_F)
\theta(k_F-h_2) 
\nonumber\\
&& \times
\delta(\np'_1+\np'_2-\nq-\nh_1-\nh_2) \nonumber
\label{hadronic12}
\end{eqnarray}
where $V/(2\pi)^3 = Z/(\frac8 3 \pi k_F^3)$ for symmetric nuclear
matter.  The integral (\ref{MEC}) in the nuclear hadronic
tensor can be reduced to seven dimensions when calculating the
response functions.  The function
$w^{\mu\nu}(\np'_1,\np'_2,\nh_1,\nh_2)$ represents the hadron tensor
for single 2p2h transitions, summed up over spin and isospin,
\begin{equation}
w^{\mu\nu} = \frac{1}{4}
\sum_{s_1s_2s'_1s'_2}
\sum_{t_1t_2t'_1t'_2}
j^{\mu}(p'_1,p'_2,h_1,h_2)^*_A  
j^{\nu}(p'_1,p¡_2,h_1,h_2)_A  
\label{elementary}
\end{equation}
where the two-body current
matrix elements is conveniently antisymetrized  
\begin{equation} \label{anti}
j^{\mu}(p'_1,p'_2,h_1,h_2)_A  
\equiv j^{\mu}(p'_1,p'_2,h_1,h_2) -
j^{\mu}(p'_1,p'_2,h_2,h_1) 
\end{equation}
and the factor $1/4$ in Eq.~(\ref{elementary}) accounts for the
 anti-symmetry of the two-body wave function with respect
to  exchange of momenta, spin and isospin quantum numbers.

Due to the intricate nature of
spin operators in the MEC, the summation over spin in Eq. 
(\ref{elementary}) is
numerically evaluated. 
Note that the single-nucleon tensor can be expanded
as a sum of direct-direct, exchange-exchange, and
direct-exchange interference terms. 
While in other models \cite{Nie11}, the
direct-exchange interference is not taken into account, 
the full contribution is included in our calculation. 

\section{Results}

In this section, we present predictions of the ESuSAM*+MEC
model for neutrino and antineutrino Charged Current (CC) quasielastic
scattering. Our analysis involves a comparison with experimental data
from MiniBooNE, T2K, and MINERvA collaborations.  A novel aspect of
our model is the inclusion, for the first time, of MEC two-nucleon
emission computed within the framework of the RMF theory for nuclear
matter. This approach accounts for dynamical effects in the MEC
responses through the relativistic effective mass and vector energy of
the nucleon.

\begin{figure}[t]
\centering
\includegraphics[width=8cm]{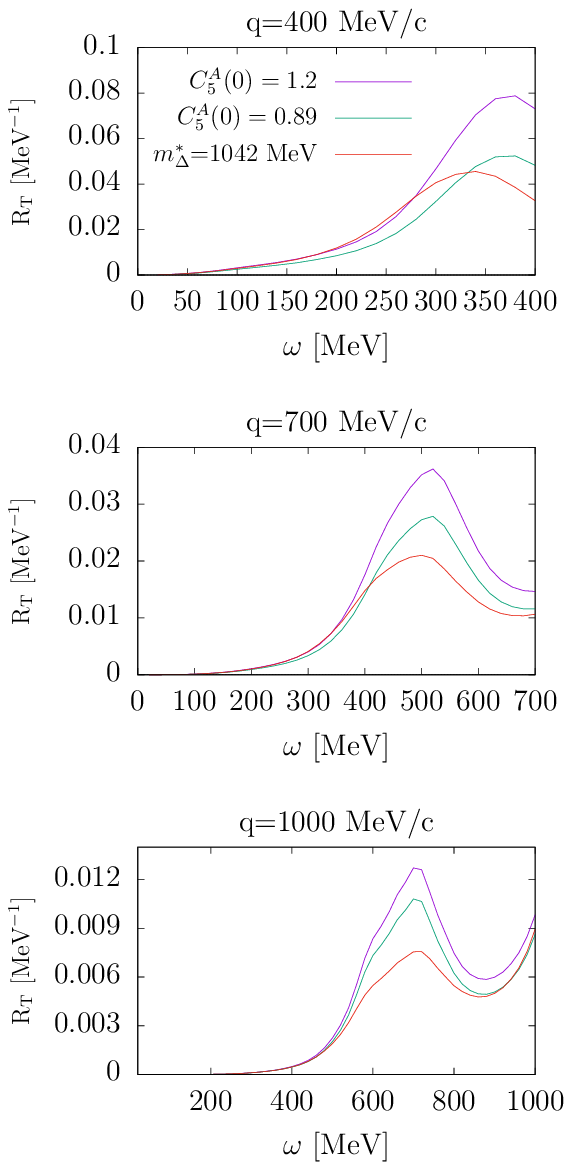}
\caption{
\deleted{{\tt cambiar labels}}
Transverse MEC response employing three distinct prescriptions for the 
$\Delta$ current: (a) \(C_5^A(0) = 1.2\), (b) \(C_5^A = 0.89\), and (c) 
\(m_\Delta^* = 1042\) MeV.
}\label{rt}
\end{figure}

While our approach shares similarities with the SuSAv2+MEC model,
which is widely used to study these reactions\cite{Meg16,Meg19}, 
it introduces significant differences in both the
treatment of scaling and the handling of MEC.

The primary difference lies in the utilization of a unified analytical
scaling function, derived from $(e,e')$ data, and the inclusion of an
averaged single-nucleon prefactor in our model. Additionally, the
scaling variable in SuSAM* incorporates the effective mass of the
nucleon. Another distinction is the decomposition of the scaling
function into two parts: a symmetric function, 
\(f^*_{1p1h}(\psi^*)\),
which represents the contribution of 1p1h processes, and a slightly
scaling-violating function 
\(f^*_{SRC}(q,\omega)\), responsible for
describing the tail of the scaling data. This latter function
phenomenologically can account for  2p2h processes generated by 
interplay between one-body currents and correlated nucleon pairs, and
interferences with MEC.
Furthermore, in our model, the treatment of MEC involves the complete
propagator of the $\Delta$ resonance, whereas the SuSAv2+MEC model only
includes the real part of the $\Delta$ propagator.

The parameters of the SuSAM* model (Fermi momentum
$k_F=225$ \added{MeV/c}, relativistic effective mass $M^*=0.8$ and  nucleon
vector energy $E_v=141$ MeV) are taken from ref. \cite{Mar21}\added{.}
The other input of the model is the
phenomenological scaling function $f_E^*(q,\omega)$, extracted from
$(e,e')$ data as described in the previous section, and in more detail 
in Ref. \cite{Mar23}.

\subsection{MEC uncertainty}

Before presenting the results for the neutrino cross sections, it is
important to address the treatment of the $\Delta$ current in the MEC
calculation. The $\Delta$ current introduces numerous uncertainties due to
its model-dependent nature. Apart from the handling of the propagator,
issues regarding form factors and the nuclear medium effect also
arise. In this article, we do not delve deeply into all these
complexities, as they are beyond the scope of this work. However, we
will briefly discuss some of them.

One primary concern pertains to the axial form factor value for the
excitation of the $\Delta$ at zero momentum transfer 
which is commonly denoted as \(C_5^A(0)\). A frequently accepted value
that has been used in many calculations is \(C_5^A(0) = 1.2\), which
we employed in our previous works. However, in Ref. \cite{Her07}, this
value was revisited through a more detailed analysis of weak pion
emission by the nucleon, and it was revised downwards to \(C_5^A(0) =
0.89\). This revised value is utilized in this work. We made this
choice because, otherwise, the MEC response would be overly enhanced
when compared with the data, as this value enters the calculations
squared. The new value yields more reasonable results, as will be
evident in the subsequent subsections.  In Figure \ref{rt}, we
contrast the computed transverse response using the two values
\(C_5^A(0)=0.89\) and 1.2. It is evident that the response
significantly increases with a larger axial form factor. Henceforth,
all results presented utilize the smaller value 0.89.

In this study, we do not account for modifications of the $\Delta$
within the nuclear medium; that is, we employ the $\Delta$ current in
vacuum. Nonetheless, we can speculate about the implications of
considering interactions between the $\Delta$ and the RMF. We suppose
that, similar to the nucleon, the $\Delta$ may acquire an effective mass
and vector energy due to these interactions. This topic was also
addressed in Reference \cite{Mar21b}, where an estimate of 
\(M_\Delta^* = 1042\) MeV/c was proposed under the assumption of
universal coupling, implying the same vector energy as the
nucleon. This value is not firmly determined.  Fig. 2 presents an
illustrative example of the transverse response when such $\Delta$-medium
interactions are taken into account. It is evident that the response
experiences a reduction compared to the case of  $\Delta$ in the vacuum.

Therefore, by examining the comparison of the transverse response in
Figure 2, which incorporates modifications to the axial form factor
and interactions with the $\Delta$ in the medium, we can gauge the
inherent uncertainty in the subsequent results regarding the MEC
effects. This insight would suggest an uncertainty band for the 2p2h
response on the order of \(\pm 20\%\) at most due to these effects.

\begin{figure*}[th]
\centering
\includegraphics[width=14cm]{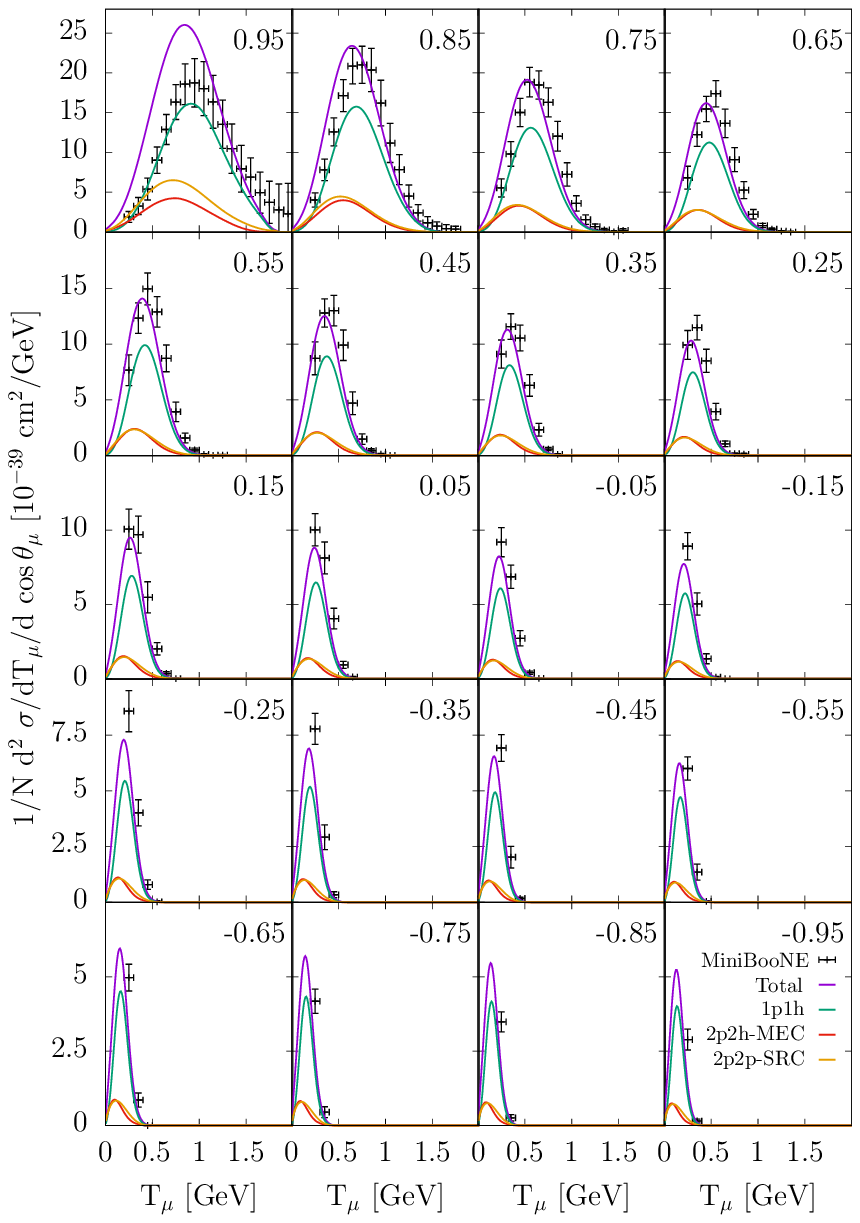}
\caption{ Flux-integrated double-differential cross section for
  neutrino scattering on $^{12}$C corresponding to the MiniBooNE
  experiment. The experimental data are from Ref.\cite{Agu10}
}\label{nu-miniboone}
\end{figure*}

\begin{figure*}[thp]
\centering
\includegraphics[width=14cm]{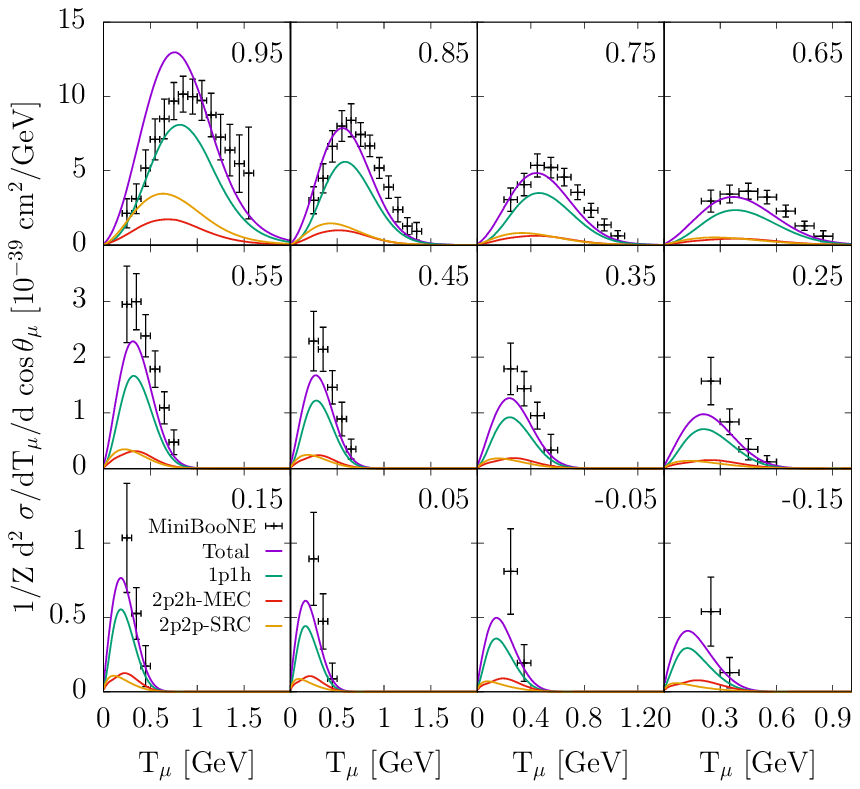}
\caption{MiniBooNE experiment flux-integrated double-differential cross
  section for antineutrino scattering on $^{12}$C.The
  experimental data are from Ref. \cite{Agu13}. }\label{antinu-miniboone}
\end{figure*}

\subsection{MiniBooNE}

Here  we present results for the kinematics of the
$(\nu_\mu,\mu^-)$ reaction on \(^{12}\)C as conducted in the MiniBooNE
experiment, depicted in Figure \ref{nu-miniboone}. In this experiment, the
flux-averaged cross-section was measured as a function of the muon
energy for  fixed values of \(\cos\theta_\mu\). The kinematics in terms of
\(T_\mu\) (muon kinetic energy) and \(\cos\theta_\mu\) (angle with respect
to the incoming neutrino direction) are averaged over bins.
The flux-averaged double differential
 cross section is computed as
\begin{equation} \label{average}
\frac{d^2\sigma}{dT_\mu d\cos\theta_\mu}
=
\frac{
\int dE_\nu \Phi(E_\nu)
\frac{d^2\sigma
}{dT_\mu d\cos\theta_\mu}(E_\nu)}
{\int dE_\nu \Phi(E_\nu)}, 
\end{equation}
where $\Phi(E_\nu)$ is the neutrino flux
and
$\frac{d^2\sigma
}{dT_\mu d\cos\theta_\mu}(E_\nu)$
is the cross section for fixed neutrino energy $E_\nu$.

In Figure \ref{nu-miniboone}, we show the individual contributions of the 1p1h
responses (\(R_{1p1h}\)), the tail responses (\(R_{SRC}\)), and the
2p2h-MEC responses (\(R_{MEC}\)), as well as the total contribution
obtained by summing these three components.

The computation of MEC responses entails intricate calculations,
involving seven-dimensional integrals of the 2p2h
responses. Additionally, there are integrations over neutrino energy
for the flux-averaged cross-section and possibly an additional
averaging over bins in \(\cos\theta_\mu\). To expedite these computations,
a parametrization of the MEC responses was proposed in \added{Ref. }
\cite{Mar21b}. This parametrization introduces semi-empirical formulas
that factorize coupling coefficients, form-factors,  phase-space, 
and the averaged $\Delta$ propagator. 
The semi-empirical MEC formulas incorporate adjustable
coefficients dependent on \(q\), which are tabulated in
\cite{Mar21b}. This semi-empirical approach proves highly
advantageous, particularly due to its flexibility, enabling the
modification of numerous model parameters without necessitating a
re-adjustment of the coefficients. This flexibility may facilitate the
study of systematic errors associated with the model parameters. In
the case of the 1p1h and SRC responses, no such challenges are
encountered, as they are represented by analytical functions.

The results presented in Figure \ref{nu-miniboone} are displayed in
panels representing different bins of \(\cos\theta_\mu\), where the
flux-averaged cross section per neutron is plotted as a function of
the muon \added{kinetic} energy $T_\mu$. The width of the experimental
bin is \(\Delta\cos\theta_\mu=0.1\).  The results are shown as an
average over the bin, taking five equidistant points within each bin.

For extremely small angles, particularly 
\(\cos\theta_\mu=0.95\), our
model clearly overestimates the data. In these cases, the transferred
momentum (\(q\)) is also very small for the neutrinos that dominate
the flux. For \(q < 200\) MeV/c, the scaling model is not suitable due
to the breakdown of the conditions required for the factorization
approximation. At such small momenta, finite-size effects are important
 and the description of nuclear final states as plane waves is not valid.
For larger scattering angles, the description of the data improves and
is generally in agreement within the bounds of experimental
error. However, a small shift of the data is observed compared to the
theoretical distribution.

The impact of the MEC is nearly identical to that of the SRC, except
once again for very small angles where the SRC effects are more
pronounced. The peaks of the MEC and SRC responses almost coincide,
with the maximum being shifted to lower energies compared to the peak
of the 1p1h response.   It is evident
that the inclusion of both 2p2h contributions plays a significant role
in accurately describing the observed data.  The total effect of 2p2h
contribution with respect to the total cross section is approximately
30\%. 

The antineutrino scattering case, as compared with MiniBooNE data,
exhibits similar trends to the neutrino case, as shown in Figure
\ref{antinu-miniboone}. \added{In this case, the cross section averaged over flux is divided by the number of protons and plotted as a function of the muon's kinetic energy. }For very small angles, the data are
overestimated, as expected, and for larger angles, the description is
reasonable, although generally, the data are underestimated. The
effects of MEC and SRC correlations are comparable, except at very
small angles where SRC dominates over MEC. However, this is again a
regime where the model's validity is questionable. The overall impact
of the 2p2h contribution amounts to approximately 30\% of the total
cross section.

We should clarify that the combined contributions referred to as 1p1h
and SRC approximately coincide with the non-extended SuSAM* model
employed in reference \cite{Rui18} for studying the same reaction. In
that model, the scaling function was not decomposed into its symmetric
and tail components. Given that both models were fitted to electron
data, it is reasonable that they yield similar results by design.  Our
extended SuSAM* model holds the advantage of enabling the separation
of the high-energy tail contribution from the scaling function, which
is presumed to originate from the distribution of high
momenta. Meanwhile, the 1p1h contribution corresponds to nucleons with
moderate momenta.

On the other hand, the contribution of the MEC in the our approach
is typically of a similar magnitude, to what was obtained in the
SuSAv2+MEC model \cite{Meg16,Meg19}. The latter model neglected the
imaginary part of the $\Delta$ propagator. In our current results, the
complete propagator is employed, which tends to increase and shift the
response to higher energies.  However, the nucleons are described
within the RMF framework, with effective mass and vector energy,
leading to a reduction in the response. As a result, the combined
effect yields responses of comparable magnitudes in both models. The
quantitative differences between the two models were studied in
Reference \cite{Mar21}.

\begin{figure}
\centering
\includegraphics[width=8cm]{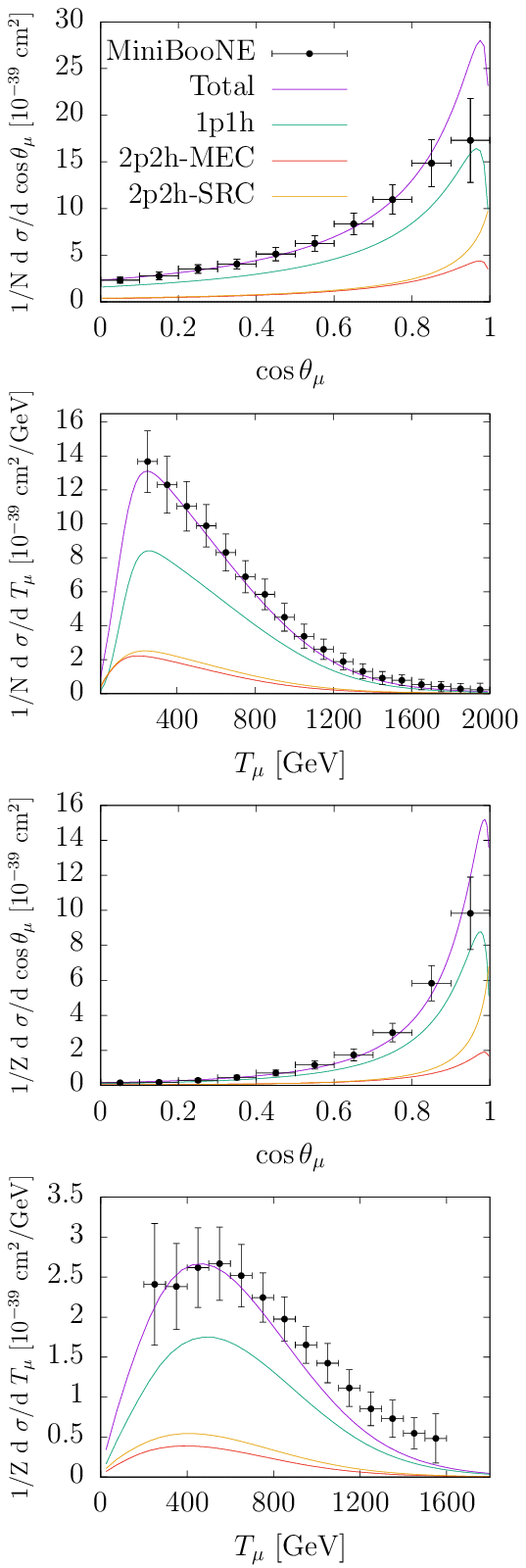}
\caption{\deleted{
{\tt arreglar ylabel}}
MiniBooNE flux-averaged CCQE single-differential cross section as a
  function of $\cos\theta_\mu$ and $T_\mu$ for neutrino (top
  panels) and antineutrino scattering (bottom panels) from $^{12}$C. The
  experimental data are from Ref. \cite{Agu10,Agu13}
}\label{single-miniboone}
\end{figure}

\begin{figure}
\centering
\includegraphics[width=8cm]{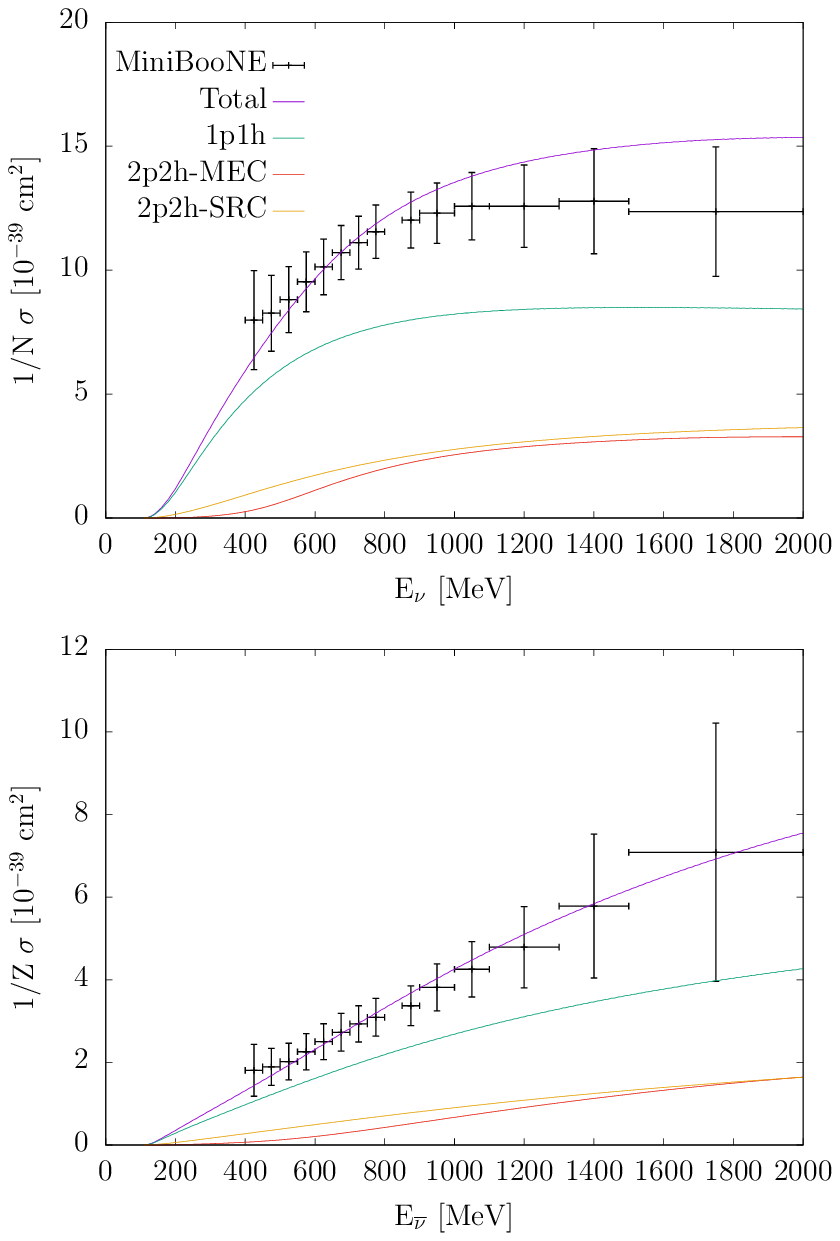}
\caption{
MiniBooNE
  CCQE total neutrino (top panel) and atineutrino (bottom panel) cross
  section on $^{12}$C.  The experimental data are from
  \cite{Agu10,Agu13} }
\label{sigma-miniboone}
\end{figure}

In Figure \ref{single-miniboone}, a more comprehensive perspective of
the previous results can be gained by integrating over 
\(\cos\theta_\mu\)
or \(T_\mu\) to obtain the single-differential cross section with
respect to \(T_\mu\) or \(\cos\theta_\mu\).

For neutrino scattering, the distribution in \(\cos\theta_\mu\) is well
reproduced except at very forward angles, while the experimental
\(T_\mu\) distribution is accurately reproduced across all values. A
similar conclusion can be drawn for antineutrino scattering, except
that the \(T_\mu\) distribution is slightly underestimated at high
energies. In both cases, the contribution of 2p2h interactions proves
essential to replicate the data.

Apart from the issues with very forward angles that have been
previously mentioned, Figure \ref{single-miniboone} highlights that
the SRC response tends to increase significantly for 
\(\cos\theta_\mu \sim
1\). This effect is attributed to the extrapolation of the 2p2h
parameter \(c^{pn}(q)\) to \(q \rightarrow 0\), while the coefficients
were fitted from \(q = 100\) MeV/c  onwards. These results can be
improved by excluding the contributions from the SRC tail for \(q\)
values less than 100 MeV/c, as their \(q\)-dependence becomes questionable
in this region.

In Figure \ref{sigma-miniboone}, we present the MiniBooNE unfolded
total cross-section per neutron (proton) as a function of
neutrino(antineutrino) energy The predictions from our comprehensive
model align well with the data, although in the case of neutrino
interaction, the prediction slightly exceeds the data, considering the
error bars.  The contributions from MEC and SRC are of comparable
magnitude, each accounting for more than 20\% of the total cross
section

Overall, our results do not significantly
deviate from those of the SuSAv2+MEC model. Therefore, the
ESuSAM* model provides an alternative  scaling plus 2p2h framework to
describe the neutrino cross-section. 
It is  remarkable that the agreement of our results  with the
MiniBooNE data is similar to that obtained with more sophisticated
models  \cite{Nie12,Nie13,Mar11,Mar13,Gal16}. 
This is so because our model incorporates
dynamic effects in the nucleon due to the RMF,
like enhancement of transverse response due to
lower components of nucleon spinors and other nuclear effects hidden
into the phenomenological scaling function.

\begin{figure*}
\centering
\includegraphics[width=16cm]{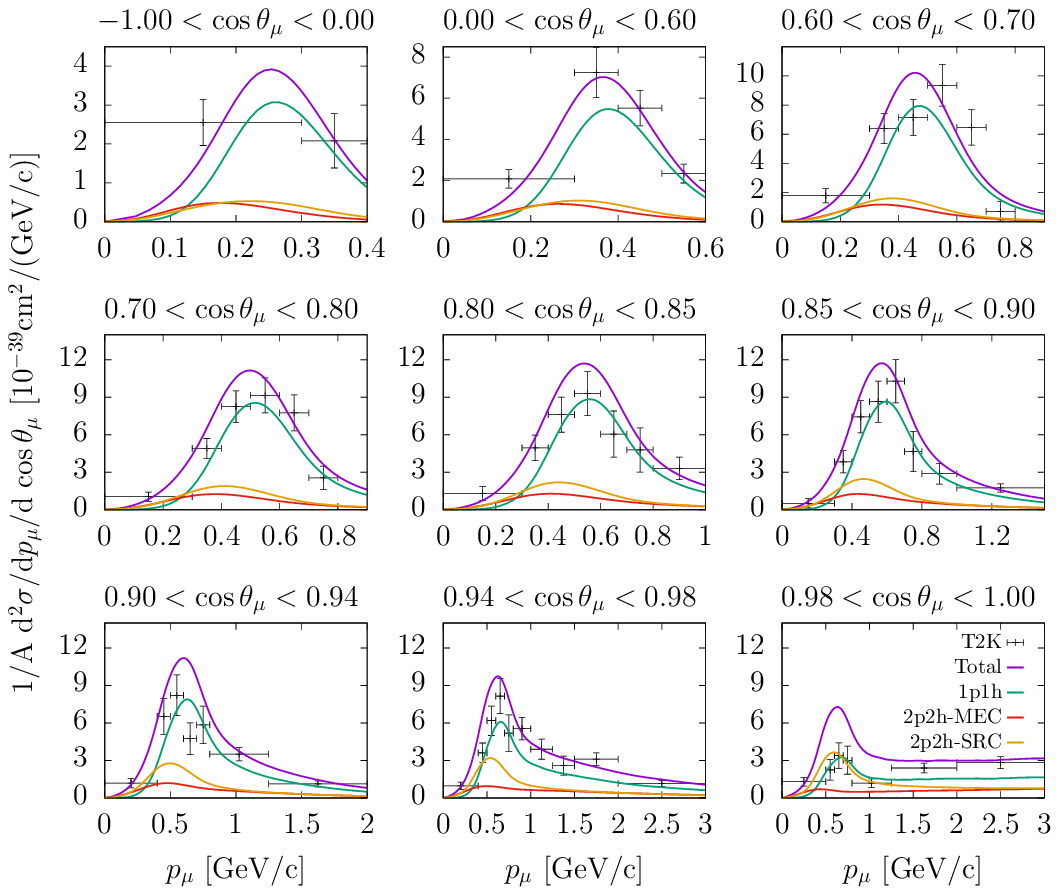}
\caption{
T2K flux-folded double differential CCQE cross section per nucleon for $\nu_\mu$
 scattering on $^{12}$C in the extended SuSAM*+MEC model.
Experimental data are from Ref. \cite{Abe16}. }
\label{t2k}
\end{figure*}

\subsection{T2K}

In Figure \ref{t2k}, we depict the flux-folded CC double-differential
cross-section for \(\nu_\mu\)-$^{12}$C scattering from the T2K experiment 
\cite{Abe16}
in comparison with
the predictions derived from the ESuSAM*+MEC model. In this
experiment, the bins in \(\cos\theta_\mu\) possess a variable size. For
larger angles, the bins are wider, while they become significantly
smaller for smaller angles. This means that, for larger angles, we
are averaging over a broader angular interval, thereby encompassing
numerous values of \(q\), which are generally large. This aligns with
the model's presumption of performing optimally at higher \(q\)
values. 

It's important to note that the neutrino fluxes for T2K and MiniBooNE
experiments are different, making the direct comparison of their
results challenging. The T2K experiment utilizes a narrower neutrino
flux centered around an energy of about 500 MeV, with an elongated
tail. In contrast, the MiniBooNE experiment employs a broader and more uniform 
flux
extending to energies of around 1.5 GeV. Consequently, different
energy averages are being considered in these two experiments.

In general, our model reasonably reproduces the data, although it
tends to slightly overestimate them as the \(\cos\theta_\mu\) bin size
increases. An exceptional case is observed in the bin \(0.98 <
\cos\theta_\mu < 1\), which corresponds to very forward angles. In this
case, the model doubles the experimental cross-section at the maximum,
which corresponds to the lowest \(q\) region where the model's
validity is questionable.  However at this kinematics the tail of the
cross section for high values of the muon momentum is well reproduced.

Furthermore, the impact of the MEC is not as pronounced as observed in
the MiniBooNE case, and its peak occurs at lower energies than the
1p1h response. This behavior is attributed to the emphasis of
different regions in the flux average, where the MEC has less
significance. The effect arising from the tail of the scaling function
is similar to that of the MEC for larger angles but becomes more
prominent as the angle decreases. It can be stated that for forward
angles, the tail effect is increasingly emphasized. This emphasis
reaches its zenith in the last bin, where the forward scattering is
most significant. In this scenario, the SRC response is of similar
magnitude as the 1p1h response. However, this finding should be
interpreted cautiously due to the intricate kinematic considerations
involved in this regime.

\begin{figure*}
\centering
\includegraphics[width=16cm]{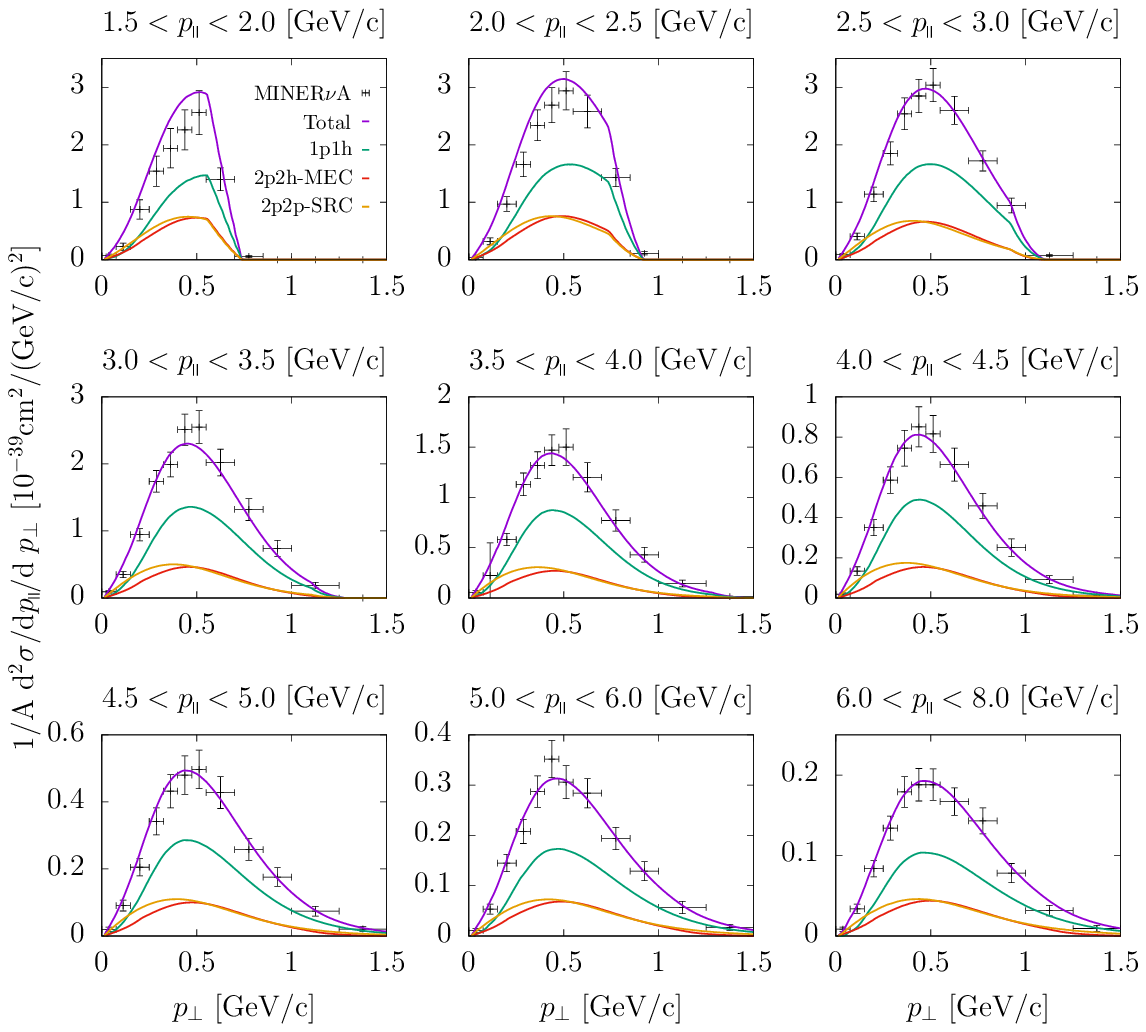}
\caption{ 
MINERvA flux-folded double-differential cross section for
  muon neutrino scattering on hydrocarbon (CH). The QE-like
  experimental data are from Ref. \cite{Rut19}.}
\label{minerva}
\end{figure*}

\begin{figure*}
\centering
\includegraphics[width=16cm]{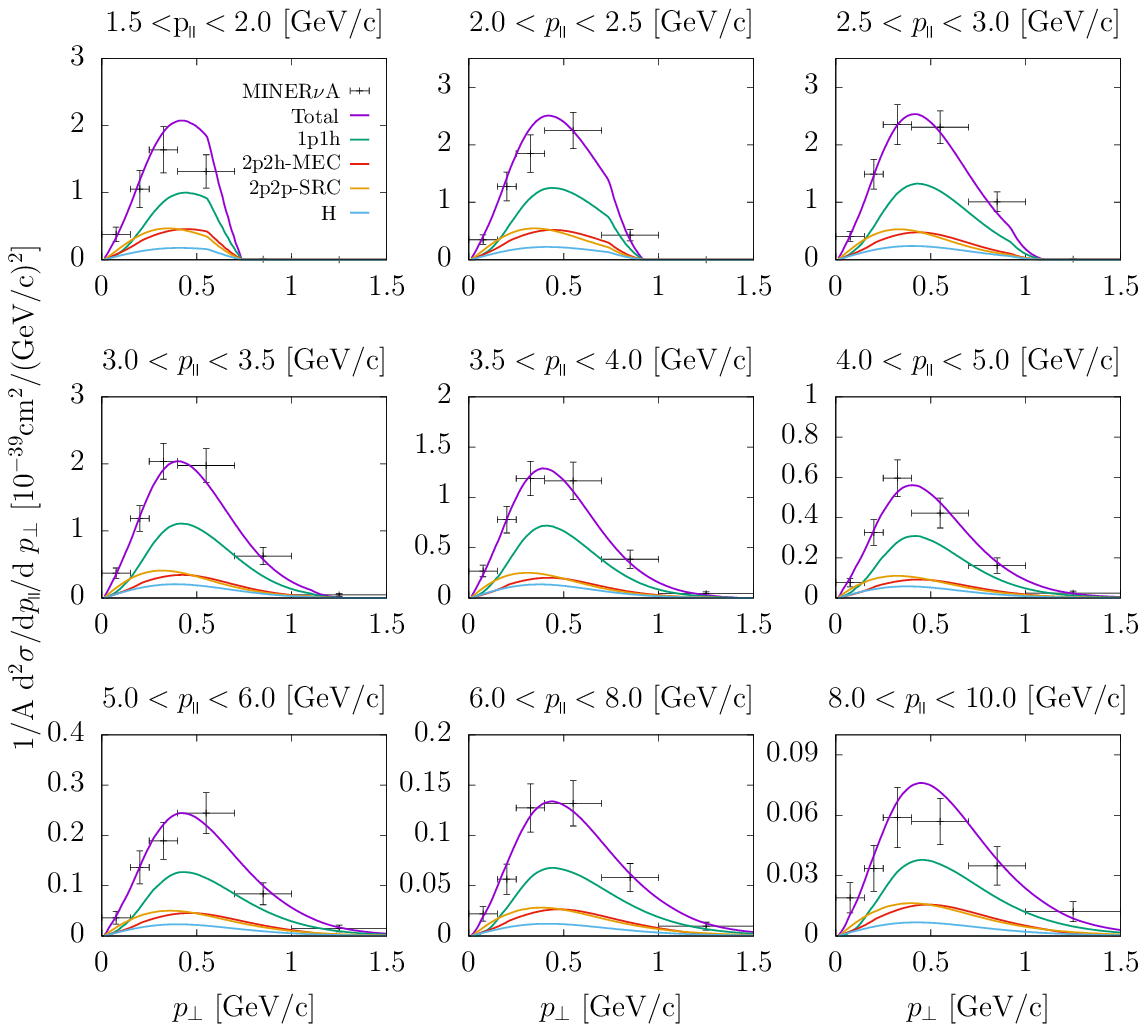}
\caption{ \deleted{{\tt los errores deben estar mal. Comparar con paper 3. 
Se han dividido dos veces por A? Los
    tres paneles inferiores subirlos hasta 0.4, 0.2 y 0.1. El key
    ponerlo en el primer panel con letra mas pequeña y con los
    siguientes keys: (a) MINERvA, (b) Total, (c) 1p1h (d) MEC, (e)
    SRC, (f) H.}}
MINERvA flux-folded double-differential cross section for
  antineutrino scattering on hydrocarbon (CH). In antineutrino
  scattering, The single-nucleon contribution for the proton in H is also shown. The QE-like experimental data are from
  Ref. \cite{Pat18}.}
\label{minerva2}
\end{figure*}

\subsection{MINERvA}

The experimental data from MINERvA are provided as a function of the
measured muon longitudinal and transverse momenta
\cite{Bet16,Pat18,Rut19,Lu21,Rut21,Kle23}.  The double-differential
cross section is expressed in terms of the longitudinal and transverse
momenta of the scattered muon, denoted as \(p_\parallel\) and
\(p_\perp\) respectively
\[
p_\parallel = p_\mu \cos \theta_\mu, \kern 1cm 
p_\perp = p_\mu \sin \theta_\mu.
\]
These expressions establish the connection between the kinematic quantities in the context of the MINERvA experimental data, and
there is the following relation between the cross sections 
in terms of both sets of variables:
\begin{equation}
\frac{d^2\sigma}{d p_{\parallel} dp_{\perp}} =
\frac{\sin\theta_\mu}{E_\mu}
\frac{d^2\sigma}{d E_\mu d\cos\theta_\mu}.
\end{equation}

In Figs. \ref{minerva} and \ref{minerva2} our results are compared
with the QE-like data collected using a hydrocarbon target (CH)
\cite{Rut19,Pat18} for neutrino and antineutrino scattering,
respectively.
The computed flux-averaged cross sections 
and the MINERvA collaboration's data are in good
agreement. Thanks to the high-energy neutrino flux, which can extend
beyond 10 GeV, the involved momentum transfers are generally
substantial, around 1 GeV/c, creating favorable kinematic conditions for
the scaling model to describe the data.

Both the 2p2h contributions, including MEC and SRC, are of comparable
magnitude, around 20\% or more, and are crucial for capturing the data
trends. The peaks of the 1p1h and 2p2h distributions practically
coincide at the same value of \(p_\perp\). The region where the
extended SuSAM* + MEC model appears to struggle more is the first
\(p_\parallel\) bin, which corresponds to the lowest momentum transfer
values, where the data are slightly overestimated for neutrino scattering on
the left side of the curve. However, across the rest of the
kinematics, the model aligns with the data within the error bars.

The strong agreement between the current model and the cross section
data from the MINERvA experiment serves as a paradigmatic example. It
demonstrates how a scaling model with 1p1h and 2p2h contributions
derived from electron data can reproduce the neutrino cross section
with only the modification of the current operator. This suggests that
the more intricate aspects of nuclear structure relevant to the
reaction are effectively incorporated implicitly within both the 1p1h
and SRC scaling functions for the experiment's kinematics.

This outcome also represents a positive validation for the 2p2h-MEC
model based on the RMF framework, which encompasses dynamic components
of nucleons within the nuclear medium. The successful agreement
further reinforces the model's capability to describe neutrino
interactions within these experimental conditions.

\section{Conclusions}

In this study, we have employed an extended superscaling analysis with
a relativistic effective mass to compute neutrino and antineutrino
cross sections. Given that a significant source of systematic error in
neutrino oscillation experiments arises from the lack of a complete
theoretical description of neutrino-nucleus interactions,
scaling-based models can offer valuable, alternative insights for such
analyses. The scaling transformation can be regarded as an approximate
symmetry of the nuclear response, yielding a universal function that
solely depends on a scaling variable in the quasielastic
regime. Changes in kinematics that keep the scaling variable invariant
yield the same scaling function. In other words, different probes with
distinct initial and final momentum transfers,  $k_i$ and $k_f$,
expanding the same scaling variable, 
but exploring different $(q,\omega)$
spectral values,  have the same scaled response.

The superscaling approach leverages the experimental scaling
information obtained from quasielastic electron scattering to predict
neutrino cross sections, assuming that the scaling function remains
relatively insensitive to the specific type of lepton-nucleus
interaction. This assumption allows for the extrapolation of the
scaling function from electron scattering to neutrino interactions,
providing a consistent framework to analyze and interpret
neutrino-nucleus cross section data. By utilizing this approach, the
theory-driven scaling function helps bridge the gap between different
experimental measurements, providing a useful tool for understanding
neutrino interactions across various experiments and energy ranges.

The extended SuSAM* approach employed in this study for (anti)neutrino
scattering introduces several novel features and modifications. The
scaling function is directly extracted from $(e,e')$  data by subtracting
the contribution of the 2p2h-MEC response and selecting quasielastic
data points that collapse, while removing those that
don't. Additionally, the phenomenological scaling function is
decomposed into a sum of a symmetric function and the tail
contribution, which is parametrized using a factorized 2p2h-like
model. 

The ESuSAM*+MEC model relies on the 
RMF model of nuclear matter, a key feature that brings in
dynamic relativistic ingredients. This sets it apart from the
traditional approaches that often use the RFG
model. 
The dynamic relativistic ingredients of the RMF, when applied to both the
scaling function and the MEC, capture aspects of nucleon interactions
that are crucial for understanding neutrino-nucleus interactions in a
broader energy and momentum range.

The  model includes three contributions: the 1p1h
response, the 2p2h-MEC response and the 2p2h-SRC response.  As a
cautionary note, while SRC certainly contribute to the emission of two
particles, the contribution referred to here as 
"2p2h-SRC" cannot be
exclusively attributed to correlations alone. This is because
correlations cannot be disentangled from other effects that also
contribute to the tail of the scaling function, such as interference
with MEC, final-state interactions, and other ingredients that violate
scaling. At best, it can be said that this contribution serves as an
upper limit for the effects of SRC on the nuclear response. In any
case, it's useful to decompose the results into the genuine
contribution from 1p1h emission and the contribution from the tail,
which is compatible with 2p2h events.  By distinguishing between these
contributions, the model provides a more nuanced understanding of the
underlying nuclear dynamics and the various factors that contribute to
the observed responses in neutrino scattering.

By utilizing this framework, we have successfully reproduced a wide
range of experimental data from MiniBooNE, T2K, and MINERvA
collaborations.  The fact that the model performs well, especially at
higher momentum transfers, indicates that it captures the underlying
physics of neutrino-nucleus interactions in those regimes. 
The challenges arise when
dealing with very low momentum transfers, where the model's
predictions deviate from the data. This is not uncommon, as the
theoretical description of very low momentum transfers can be complex
due to effects like nuclear shell structure and the limitations of the
model assumptions. Using a shell model approach for
low $q$ is more reasonable, as shell models are better suited for
describing the behavior of nucleons in those energy regimes.

The significant contribution of the 2p2h response to the overall
response, particularly for experiments with higher momentum transfers
like MINERvA, underscores the importance of considering multi-nucleon
emission in neutrino-nucleus interactions. The comparable impact
of MEC and SRC effects further emphasizes their significant roles in
the response, which is consistent with the complexity of these
interactions.

This extension of the SuSAM*
model, along with its successful comparison to data, not only provides
valuable insights into neutrino interactions but also underlines the
significance of considering dynamic nucleon effects within the nuclear
medium. These findings emphasize the potential of scaling-based models
to contribute to a deeper understanding of neutrino-nucleus
interactions and their implications for neutrino oscillation
experiments.

\section{Acknowledgments}

Work supported by:
Grant PID2020-114767GB-I00
funded by MCIN/AEI/10.13039/501100011033; FEDER/Junta de
Andalucia-Consejeria de Transformacion Economica, Industria,
Conocimiento y Universidades/A-FQM-390-UGR20; Junta de
Andalucia (Grant No. FQM-225).

\end{document}